\documentstyle[12pt]{article}
\newcommand{\xp}{\vec{x}_P}
\newcommand{\xd}{\vec{x}_D}

\newcommand{\vq}{\vec{q}}
\newcommand{\vL}{\vec{L}}
\newcommand{\vn}{\vec{n}}
\newcommand{\ve}{\vec{e}_z}
\newcommand{\sa}{\sqrt{A}}
\newcommand{\e}{\mbox{e}}
\def\greaterthansquiggle{\raise.3ex\hbox%
                        {$>$\kern-.75em\lower1ex\hbox{$\sim$}}}
\def\lessthansquiggle{\raise.3ex\hbox{$<$\kern-.75em\lower1ex\hbox{$\sim$}}}
\newcommand{\be}{\begin{equation}}
\newcommand{\ee}{\end{equation}}
\newcommand{\ba}{\begin{eqnarray}}
\newcommand{\ea}{\end{eqnarray}}
\newcommand{\no}{\nonumber}

\newcommand{\cL}{{\cal L}}

\normalsize
\begin{document}
\bibliographystyle{plain}
\begin{titlepage}
\begin{flushright}
UWThPh-1996-17\\
\today\\
\end{flushright}
\vspace{2cm}
\begin{center}
{\Large \bf  Real Oscillations of Virtual Neutrinos}\\[70pt]

W. Grimus and P. Stockinger\\
Institut f\"ur Theoretische Physik\\
Universit\"at Wien\\
Boltzmanngasse 5\\
A-1090 Vienna, Austria

\vspace{5cm}

{\bf Abstract}\\[7pt]
\end{center}

We study the conditions for neutrino oscillations in a field
theoretical approach by taking into account that only the
neutrino production and detection processes,
which are localized in space around the coordinates
$\xp$ and $\xd$, respectively, can be manipulated.
In this sense the neutrinos whose oscillations are investigated
appear as virtual lines connecting production with detection in
the total Feynman graph and all neutrino fields or states to be
found in the discussion are mass eigenfields or eigenstates.
We perform a thorough examination of
the integral over the spatial components of the inner neutrino
momentum and show that in the asymptotic limit $L=|\xd - \xp|
\rightarrow \infty$ the virtual neutrinos become ``real'' and
under certain conditions the usual picture of neutrino
oscillations emerges without ambiguities.
\end{titlepage}

\section{Introduction}

In particle physics, oscillation phenomena have so far only been
observed in the $K^0 \bar{K}^0$ and $B^0 \bar{B}^0$ systems.
Analogous oscillations are expected to take place between
neutrinos of different lepton flavours  \cite{pon} if neutrinos are massive
particles and if individual lepton numbers are not conserved.
This non--conservation manifests itself in a neutrino mass
matrix which is non--diagonal in the weak basis, i.e. the
basis where the mass matrix of the charged leptons is diagonal
and the charged current interactions of the standard model (SM)
is of the form
\be \label{cc}
\cL_{CC} = - \frac{g}{\sqrt{2}} \left( \sum_{\ell = e,\mu,\tau}
\bar{\nu}_{\ell L} \gamma^\mu \ell_L W^+_\mu + \mbox{h.c.} \right).
\ee
Though neutrino masses and mixings rank among physics beyond the
SM it is nevertheless reasonable to assume that neutrino
production and detection are within the province of the SM
interactions. In this sense the left--handed neutrino fields in
eq. (\ref{cc}) are linear combinations of mass eigenfields, i.e.
\be \label{mix}
\nu_{\ell L}(x) = U_{\ell j} \nu_{jL}(x)
\ee
with a unitary $3 \times 3$ matrix $U$ and neutrino fields
$\nu_j$ with mass $m_j$.

The standard approach to neutrino oscillations (see, e.g., ref.
\cite{bil}) is important for its
simplicity and its physical insight but is not sufficient for a complete
understanding. (See, e.g., ref. \cite{rich1} for a list of its
conceptual problems.) It is well known that it works only in the
extreme relativistic limit. Wave--packet treatments \cite{kay} are not
totally satisfactory either without knowing size and form of the
wave packets. However, it is clear that, if the particles involved
in the neutrino production process are assumed to have definite
4--momenta, the neutrino is forced to have a definite
4--momentum too and hence is in one of the mass eigenstates.
Therefore, in order to observe neutrino oscillations one needs a
sufficient spread of momentum (or energy) of the particles 
in the production and detection mechanism. It has been stressed in the
literature \cite{rich1,giunti}
that the correct way to avoid all difficulties and
ambiguities associated with neutrino oscillations is to
concentrate on those things which can really be manipulated or
observed like the particles responsible for neutrino production
and the target responsible for neutrino detection. This amounts
to including the production and detection mechanisms
localized at the coordinates $\xp$ and $\xd$, respectively, in the
calculation such that the neutrinos propagating from $\xp$ to
$\xd$ are virtual particles \cite{rich2} or intermediate
states (see ref. \cite{giunti}
for a field--theoretical treatment and ref. \cite{rich1} for a
quantum--mechanical treatment along these lines).

In this paper we choose the field--theoretical approach and
apply it to the antineutrino production as it occurs in a reactor and
assume that the antineutrinos are detected by elastic electron scattering.
(This is, e.g., the situation in the MUNU experiment \cite{munu}.) The
antineutrino flux is produced by the $\beta$--decay of the
fission products in the reactor. We simplify this mechanism by taking only the
$\beta$--decay of the neutron and assuming that both neutron and
proton are bound to nuclei at rest. This merely
serves to keep our notation as simple as possible and has no
impact on the essence of our discussion. The main points in our
treatment of neutrino oscillations are the following:
\begin{itemize}
\item The neutron and the proton are bound to nuclei in
stationary states localized at the coordinates $\xp$ whereas the
target electron is bound in an atom at the point $\xd$. Neutrino
production and detection are macroscopically separated by a
distance $L \equiv |\vec{L}|$ with $\vL \equiv \xd - \xp$.
\item All particles in the final state are described by plane
waves. This is justified for the electron in the detector and
all the more for the other electron from the $\beta$--decay and
the antineutrino which are both not observed. In the
event rate it has to be summed (integrated) over a complete set
of states for these unobserved particles and plane waves
obviously serve this purpose. (This is one of the points which
distinguishes our work from ref. \cite{giunti}.)
\item The internal (anti)neutrino is represented by its Feynman
propagator. The integration over this internal line with momentum
$q$ is coupled to the macroscopic distance between
neutrino production and detection by $\int d^3 q \exp (-i \vec{q}
\cdot \vec{L}) (q^2 -m^2_j+i\epsilon )^{-1} \ldots$ for a neutrino
with mass $m_j$. We use a mathematical
theorem which is proved in the appendix
to obtain the leading term of the amplitude in the limit
$L \rightarrow \infty$. We will see that in this asymptotic limit the
antineutrino is ``forced'' to be on its mass shell and the
momentum $\vq$ is aligned with $-\vL$.
\end{itemize}

\section{The oscillation amplitude}

We consider the following weak process
\ba \label{process}
n\rightarrow p+e^- +\bar\nu & & \no \\
& \searrow & \no \\
& & \bar\nu +e^-_D\rightarrow\bar\nu +e^-_D
\ea
occurring through the intermediate propagation of an antineutrino, where
n (neutron) and $e^-_D$ (electron) are the initial particles. As already
mentioned we assume that the
neutron, the proton and the target electron are in stationary states.
Since neutrino production and detection are localized at $\xp$
and $\xd$, respectively, the spinors of the initial particles
and the proton can be written in position space as
\ba \label{spin}
\psi_p(x)    &=& \psi_p(\vec{x}-\vec{x}_P) \exp(-iE_p t),\nonumber\\
\psi_n(x)    &=& \psi_n (\vec{x}-\vec{x}_P) \exp(-iE_n t),\nonumber\\
\psi_{eD}(x) &=& \psi_{eD}(\vec{x}-\vec{x}_D) \exp(-iE_{eD}t),
\ea
respectively. The functions $\psi_{p,n,eD}(\vec{y})$ are peaked at 
$\vec{y}=0$. The final particles will be described by plane waves.

The weak interaction Lagrangians relevant for production and
detection are
\ba \label{Lag}
{\cal L}_P(x)&=&-\frac{G_F}{\sqrt{2}} \cos\vartheta_c
\sum_j
      U_{ej}\;
      \bar e(x)\gamma^\lambda(1-\gamma_5)\nu_j(x) \, \bar p(x)\gamma_\lambda
      (1-\tilde g\gamma_5)n(x)+\mbox{h.c.}, \nonumber\\
      {\cal L}_D(x)&=&-\frac{G_F}{\sqrt{2}} \sum_{j,k} 
      \Big\{ \bar e(x) \gamma^\mu
      P_L e(x) \, \left( 2 U^*_{ej}U_{ek}+\delta_{jk}(g_V+g_A)
      \right) + \no \\
      && + \bar e(x)\gamma^\mu P_R e(x) \,
      \delta_{jk} (g_V-g_A) \Big\} \,
      \bar \nu_j(x) \gamma_\mu (1-\gamma_5)\nu_k(x),
\ea
respectively, where $G_F$ is the Fermi coupling constant, $\vartheta_c$ 
the Cabibbo angle,
$g_V=2\sin^2\theta_W-1/2$, $g_A=-1/2$, $\theta_W$ the
Weinberg angle, $\tilde g \approx 1.26$ and $P_{R,L}=(1\pm \gamma_5)/2$.
All fields occurring in eq. (\ref{Lag}) are mass eigenfields.
The amplitude for the process (\ref{process}) with an
antineutrino of mass $m_k$ in the final state is given by
\be
{\cal A}_k = 
\langle p,\bar\nu_k(\vec{p}\,'\!\!_\nu), e^-(\vec{p}\,'\!\!_e),
e^-_D(\vec{p}\,'\!\!_{eD}) | T [\int d^4x_1
              \int d^4x_2 \, {\cal L}_P (x_1){\cal L}_D(x_2)] |
                n, e^-_D \rangle \;.
\ee
With the neutrino propagators of the mass eigenstate neutrinos
\be
\langle 0 | T [\nu_j(x_1) \bar\nu_j(x_2)] | 0 \rangle = i \int
       \frac{d^4q}{(2\pi)^4}\frac{\not\! q+m_j}{q^2-m_j^2+i\epsilon}
       \e^{-iq\cdot(x_1-x_2)} \;,
\ee
we obtain the amplitude
\ba \label{amp}
{\cal A}_k & = & \frac{G^2_F \cos\vartheta_c}{2} i \sum_j
      \int d^4 x_1 \int d^4 x_2 \int \frac{d^4 q}{(2\pi)^4} 
      \e^{-iq\cdot(x_1-x_2)} \no \\
      & & \exp \left( -i(E_n-E_p)t_1-iE_{eD}t_2 \right) \,
      \exp \left( ip'_e \cdot x_1+i(p'_\nu+p'_{eD})
      \cdot x_2 \right) \no \\
      & & J_\lambda(\vec{x}_1-\vec{x}_P) \, 
      \bar u_e(\vec{p}\,'\!\!_e)\gamma^\lambda(1-\gamma_5)
      U_{ej}\biggl(\frac{\not\!q+m_j}{q^2-m_j^2+i\epsilon}\biggr)
      \gamma^\mu (1-\gamma_5) v_k(\vec{p}\,'\!\!_\nu)  \no \\
      & &\bar u_e(\vec{p}\,'\!\!_{eD})\gamma_\mu \left[ 
      P_L \left( 2U^*_{ej}U_{ek}+\delta_{jk}(g_V+g_A) \right)
      +P_R\delta_{jk}(g_V-g_A) \right] \psi_{eD}(\vec{x}_2-\vec{x}_D)
\ea
where
\be
J_\lambda(\vec{x}_1-\vec{x}_P)\equiv\bar\psi_p(\vec{x}_1-
\vec{x}_P)\gamma_\lambda(1-\tilde g\gamma_5)
\psi_n(\vec{x}_1-\vec{x}_P)\;.
\ee
Note that the amplitude (\ref{amp}) is not antisymmetric with respect
to the exchange of the final electrons because the
electron generated in the $\beta$--decay never reaches the
detector or, in other words, it is absorbed in the reactor and
thus takes part in the formation of final states orthogonal to
single electron states.

The integration over $t_1$ and $t_2$ in eq. (\ref{amp}) 
can easily be carried out
leading to the delta functions $\delta(q_0+E_2)$ and $\delta(q_0+E_1)$.
Furthermore, we use the relation
\be
\int\limits_{-\infty}^\infty dx \, \e^{-ikx} f(x+b) =
      \e^{ikb}\,\tilde f(k) \;,
\ee
where $\tilde f$ is the Fourier transform of f, in the integrations 
over $\vec{x}_1$ and $\vec{x}_2$. Thus we obtain
\ba \label{amp1}
{\cal A}_k & = & i\frac{G^2\cos\vartheta_c}{2(2\pi)^2}  
\, \e^{-i\vec{p}_1\cdot\xp-i\vec{p}_2\cdot\xd}
\sum_j \int d^4
q \, \delta(q_0+E_1)\delta(q_0+E_2) \, \e^{-i\vq \cdot \vL} \nonumber\\
      & & \tilde J_\lambda(\vec{p}_1-\vec{q}) \, 
      \bar u_e(\vec{p}_1)\gamma^\lambda
      (1-\gamma_5) U_{ej}\biggl(\frac{\not\!q+m_j}{q^2-m_j^2+i\epsilon}\biggr)
      \gamma_\mu(1-\gamma_5) v_k(\vec{p}\,'\!\!_\nu)\nonumber\\
      & & \bar u_e(\vec{p}\,'\!\!_{eD}) \gamma^\mu \left[
      P_L \left( 2U^*_{ej}U_{ek}+\delta_{jk}
      (g_V+g_A) \right) + P_R\delta_{jk}(g_V-g_A) \right] 
      \tilde\psi_{eD}(\vec{p}_2 + \vec{q})
\ea
where we have defined
\be\begin{array}{ll} E_1 \equiv E_n-E_p-E'_e, &
       \vec{p}_1\equiv \vec{p}\,'\!\!_e, \\
       E_2 \equiv E'_\nu +E'_{eD}-E_{eD}, &
       \vec{p}_2\equiv \vec{p}\,'\!\!_{eD}+\vec{p}\,'\!\!_\nu \,.
\end{array}\ee

Now only the integration over q remains.
Since our initial and final states are all energy eigenstates,
the integration over $q^0$ leads to the usual $\delta$--function
$\delta(E_1-E_2)$ expressing that the initial energy is equal to
the final energy of the total process (\ref{process}). The
analogous $\delta$--functions corresponding to momentum are
smeared out by the initial momentum distributions and therefore
we have a non--trivial $d^3q$ integration in the amplitude. We
can, however, take advantage of the fact that this integration
actually amounts to calculating a Fourier transform
and evaluating it at the coordinate $\vL$, the
macroscopic distance between neutrino production and detection
(see eqs. (\ref{spin})).
Hence we can apply the theorem proved in the appendix (see eqs.
(A.1) and (A.2))
which enables us to calculate the leading term of the amplitude for large $L$.

Let us for a moment dwell upon the mathematical requirements of this theorem
in the context of the phyiscal problem under discussion. The
wave functions eq. (\ref{spin}) represent bound states. It is
therefore reasonable to assume that they fall off exponentially
and consequently the Fourier transforms $\tilde J_\lambda$ and
$\tilde \psi_{eD}$ are infinitely many times differentiable.
This meets the requirement of differentiability in the theorem.
It remains to discuss the behaviour of $\tilde J_\lambda$ and 
$\tilde \psi_{eD}$ for $|\vq| \rightarrow \infty$. If bound
states in coordinate space are also infinitely many times
differentiable in addition to their exponential fall--off then
their Fourier transforms have the same properties. However, 
bound state wave functions are in general
not everywhere infinitely many times differentiable in {\em coordinate
space}. A typical example is given by the ground state wave 
function of the hydrogen atom which is not differentiable at the
origin. Its Fourier transform nevertheless decreases with the
fourth power of $|\vq|$ which meets the conditions for the
decrease of the function in the theorem. Therefore the physics under
consideration seems to comply naturally with the mathematical
requirements.

Noting that the constant $A$ of the theorem corresponds to $q_0^2-m_j^2$,
eq. (A.2) tells us that
the asymptotic limit $L \to \infty$ of eq. (\ref{amp1}) is given
by
\ba \label{ampa}
      {\cal A}^\infty_k & = & -i \, \frac{G^2_F \cos\vartheta_c}{4}
      \, \frac{1}{L} \, \delta(E_1-E_2) \, 
      \e^{-i\vec{p}_1\cdot\vec{x}_P-i\vec{p}_2\cdot\vec{x}_D} \no \\
      & &  \sum_j \e^{iq_jL}  \tilde J_\lambda(\vec{p}_1+q_j\vec{l}\,) \, 
      \bar u_e(\vec{p}_1)
      \gamma^\lambda(1-\gamma_5) U_{ej} (q_0\gamma^0+q_j\vec{l}\!\cdot\!
      \vec{\gamma}+m_j)\gamma_\mu(1-\gamma_5)v_k(\vec{p}\,'\!\!_\nu) \no \\
      & & \bar u_e(\vec{p}\,'\!\!_{eD})\gamma^\mu \left[ P_L
      \left( 2U^*_{ej}U_{ek}+ \delta_{jk}
      (g_V+g_A) \right) + P_R\delta_{jk}(g_V-g_A) \right] 
      \tilde \psi_{eD}(\vec{p}_2-q_j \vec{l}\,)
\ea
with
$$
\vec{l}\equiv \frac{\vec{L}}{|\vec{L}|}, \quad
q_0 = -E_1=-E_2 \quad \mbox{and} \quad q_j \equiv \sqrt{q_0^2-m_j^2} \,.
$$
This equation can be interpreted in such a way that ${\cal
A}^\infty_k$ is a sum over terms which contain a
real antineutrino of mass $m_j$. The reason is that the 4--vector
$p_j$ defined by
\be
p_j^0 \equiv E_\nu \equiv -q^0 = E_2 = E'_{eD}+E'_\nu-E_{eD}
\geq 0 \quad \mbox{and} \quad \vec{p}_j \equiv q_j \vec{l}
\ee
can be identified as the 4--momentum of this antineutrino. The
spatial part of $p_j$ correctly points from $\xp$ to $\xd$. Furthermore,
with a suitable renormalization of the 4--spinors $v$ the identity
\be
q_0 \gamma^0 + q_j \vec{l} \cdot \vec{\gamma} + m_j = -\not\!p_j+m_j = 
-\sum_{\pm s}v_j(\vec{p}_j,s)\bar v_j(\vec{p}_j,s)
\ee
shows that apart from the factors $1/L$ and $\exp(iq_jL)$ the
amplitude ${\cal A}^\infty_k$ is just the sum over the products
of production and detection amplitude of antineutrinos with mass
$m_j$.

\section{Oscillating amplitude and cross section}

We expect that under certain conditions the cross section for the 
reaction (\ref{process}) ($n+e_D^-\rightarrow e^-+e^-_D+\bar\nu$) exhibits an 
oscillatory dependence on the distance $L$.
Considering first the asymptotic amplitude ${\cal A}^\infty_k$ and assuming
that $E_\nu \gg m_j \, \forall j$ we can make the following remarks and
observations to this effect. (Most of them have already been
discussed in the context of the quantum--mechanical approach of
ref. \cite{rich1}.)
\begin{itemize}
\item In order to obtain ${\cal A}_j^\infty$ we have performed the
asymptotic limit $L\rightarrow\infty$. Looking at theorem proved
in the appendix it is evident
that ``$L$ large'' means $\bar E_\nu L/\hbar c \approx \bar
E_\nu L/ 2 \cdot 10^{-13}
\mbox{ MeV} \cdot \mbox{m}\gg 1$ where $\bar E_\nu$ is an average 
antineutrino energy
and $L$ is measured in meters. For every thinkable neutrino experiment this
is very well fulfilled and corrections to ${\cal A}_k^\infty$ are suppressed 
by $(\bar E_\nu L/\hbar c)^{-1/2}$.
\item  The factor $1/L$ in the asymptotic amplitude corresponds to the
geometrical decrease of the neutrino flux by $1/L^2$ in the cross section.
\item Looking at eq. (\ref{ampa}) we conclude that neutrino oscillations with
masses $m_j,m_k$ can only take place if
$$
|q_j-q_k| \lessthansquiggle \sigma_J \quad \mbox{and} \quad
|q_j-q_k| \lessthansquiggle \sigma_D
$$
where $\sigma_J$ and $\sigma_D$ are the widths of the functions $\tilde J$ and
$\tilde \psi_{eD}$, respectively. In coordinate space this simply means that
the widths of $J$ and $\psi_{eD}$ must both be smaller than the oscillation
length $l_{\mbox{{\scriptsize osc}}}= 4\pi E_\nu/|m_j^2-m_k^2|
\approx 2.48 \, 
\mbox{m} \cdot E_\nu(\mbox{MeV})/ \Delta m^2 (\mbox{eV}^2)$.
\item To evaluate this condition for reactor neutrinos we note that $\hbar c
/\sigma_J\greaterthansquiggle 10^{-15}\mbox{m}$ and $\hbar c/\sigma_D
\greaterthansquiggle 10^{-10}\mbox{m}$. From the latter estimate one gets the
condition $\Delta m^2\lessthansquiggle (100 \, \mbox{keV})^2 $ \cite{rich1}.
\item The coherence length is infinite in our calculation since all initial
and final states are stationary.
\item Our calculation was based on the assumption that neutrinos are of Dirac
nature. In the case of Majorana neutrinos the diagonal vector
currents in ${\cal L}_D$ eq. (\ref{Lag}) are identically zero. As
a consequence, in ${\cal A}_k$ eq. (\ref{amp}) the second 
$1-\gamma_5$ has to be replaced by $-2\gamma_5$ for $j=k$.
This introduces the well known difference
of order $m_j/E_\nu$ between the amplitudes of Majorana and Dirac neutrinos
\cite{bil}.
\end{itemize}

Finally we come to the cross section or event rate in order to make contact
with the usual oscillation picture. If $|q_j-q_k| \ll 
\min(\sigma_J,\sigma_D)$ is fulfilled $\forall \, j,k$ in
addition to the assumption that all
neutrinos are ultrarelativistic, then
we can take the limit $m_j \rightarrow 0 \, \forall \, j$
in all terms of ${\cal A}_k^\infty$ except the exponential factors
$\exp(iq_jL)$. Then calculating the cross section one would arrive at the same
result as can be obtained by the following heuristic consideration:
The probability to find a neutrino $\bar\nu_\ell$ at 
a distance $L$ from the source
is given by $P_\ell=|\sum_j U_{ej}U_{\ell j}^* \e^{iq_jL}|^2$. 
Thus the number of
events in elastic $\bar\nu e^-$ scattering at a certain neutrino energy 
$E_\nu$ is proportional to $\sum_\ell P_\ell \sigma(\bar\nu_\ell
e^-;E_\nu)$ where the
$\sigma(\bar\nu_\ell e^-;E_\nu) \, (\ell=e, \mu, \tau)$ 
are the elastic scattering cross sections as
given by the SM. Note that $E_\nu$ can be reconstructed via
\be
E_\nu = m_e \left( \sqrt{1+\frac{2 m_e}{T}} \, \cos \alpha-1 \right)^{-1}
\ee
where $T=E'_{eD}-m_e$ and $\alpha$ is the angle between $\vL$ and
$\vec{p}\,'\!\!_{eD}$, the momentum of the recoil electron.
Thus the event rate is given by
\be
\frac{dN}{dE_\nu} = N_0 \Big\{ \Big| \sum_j | U_{ej} |^2 \e^{iq_jL}
\Big|^2 \left( \sigma(\bar \nu_e e^-;E_\nu) - \sigma(\bar \nu_\mu e^-;E_\nu)
\right) + \sigma(\bar \nu_\mu e^-;E_\nu) \Big\}
\ee
with a normalization constant $N_0$.

In conclusion we want to stress once more the central importance
of the theorem demonstrated in the appendix. It is not only relevant 
in the context
discussed here but it shows in general under which circumstances
the total amplitude for particle production and its subsequent
scattering factorizes into a product (or sum of products) of
production amplitude and scattering amplitude. In the case of
neutrino oscillations it leads to a transparent and simple
field--theoretical treatment without ambiguities or conceptional
problems \cite{rich1}: all neutrino states or fields are mass
eigenstates or eigenfields and the neutrino ``wave packet'' is totally
determined by the production mechanism. The method described
here also provides the correct answer in cases where the
standard approach to neutrino oscillations fails. We want to
mention, however, that some more work has to be done to extend 
our method because it is not straightforwardly
transferable to the case of accelerator neutrinos and it has the
deficiency of an unrealistic infinite coherence length for 
neutrino oscillations.

\newpage
\section*{Appendix}
\setcounter{equation}{0}
\renewcommand{\theequation}{A.\arabic{equation}}
\paragraph{Theorem:} Let $\Phi : {\bf R}^3 \rightarrow {\bf
R}^3$ be a three times continuously differentiable function such
that $\Phi$ itself and all its first and second derivatives decrease
at least like $1/\vq^{\, 2}$ for $|\vq| \rightarrow \infty$,
$A$ a real number and
\be
J(\vL ) \equiv \int d^3q \, \Phi (\vq ) \, \e^{-i\vq \cdot \vL}
\frac{1}{A - \vq^{\, 2} + i\epsilon}.
\ee
Then in the asymptotic limit $L = |\vL | \rightarrow \infty$
one obtains for $A>0$
\be \label{asympt}
J(\vec{L}) = - \frac{2\pi^2}{L} \Phi(-\sqrt{A} \vec{L}/L)
\, \e^{i \sqrt{A} L} + {\cal O}(L^{-3/2})
\ee
whereas for $A<0$ the integral decreases like $L^{-2}$.\\
Remark: In order to make the proof of this theorem
transparent we will first introduce three lemmas and
then divide the proof into several steps.

\paragraph{Lemma 1:} Let $f$ be a three times continuously differentiable
function and
\be \label{I}
I(r) \equiv \int_0^\pi d\theta \sin \theta \, \e^{-ir\cos
\theta} f(\theta).
\ee
Then in the asymptotic limit $r \rightarrow \infty$ one obtains
\be
I(r)=-\frac{\e^{-ir} f(0) - \e^{ir} f(\pi)}{ir} +
{\cal O}(r^{-3/2}).
\ee
\paragraph{Proof:} We first perform a partial integration
resulting in
\be \label{part}
I(r)=-\frac{\e^{-ir} f(0) - \e^{ir} f(\pi)}{ir} - \frac{1}{ir} I_1(r)
\ee
with
\be
I_1(r)=\int_0^\pi d\theta \, \e^{-ir\cos \theta} f'(\theta ).
\ee
It is convenient to split $I_1(r)$ into the two parts
\be
I_1(r)=\int_0^{\pi/2} d\theta \, \e^{-ir\cos \theta} f'(\theta )+
\int_0^{\pi/2} d\theta \, \e^{ir\cos \theta} f'(\pi - \theta ).
\ee
In the following we only discuss the first integral in this
equation since the second one is treated analogously. Defining
$g(\theta ) =(f'(\theta ) - f'(0))/\sin \theta $ we obtain
\be \label{I1}
\int_0^{\pi/2} d\theta \, \e^{-ir\cos \theta} f'(\theta )=
\int_0^{\pi/2} d\theta \, \e^{-ir\cos \theta} f'(0)+
\int_0^{\pi/2} d\theta \, \sin \theta \, \e^{-ir\cos \theta} g(\theta ).
\ee
It is easy to show that $g$ is twice continuously
differentiable in the intervall $[0,\pi /2]$ as a consequence of
the three times continuous differentiability of $f$. Therefore one
can apply a partial integration to the second integral on the
right--hand side of eq. (\ref{I1}) in the same way as to eq.
(\ref{I}) and thus this integral decreases like $1/r$ in
the asymptotic limit (see eq. \ref{part}).
Together with the $1/r$ in front of
$I_1(r)$ in eq. (\ref{part}) there is an overall decrease of
$1/r^2$. Therefore it remains to consider the asymptotic
behaviour of
\be \label{intcos}
\int_0^{\pi /2} d \theta \, \e^{-ir \cos \theta } =
\frac{1}{2} \left( \int_0^1 dz \frac{\e^{-irz}}{\sqrt{1-z^2}} -
\int_{\pi /r}^{1+\pi /r} dz \frac{\e^{-irz}}{\sqrt{1-(z-\frac{\pi}{r})^2}}
\right).
\ee
The right--hand side of this equation allows to deduce the upper bound
\ba
\lefteqn{| \int_0^{\pi /2} d\theta \, \e^{-ir\cos \theta }| \leq}
\no \\
 & &\leq \frac{1}{2} \left[ \int_0^{\pi /r} dz \frac{1}{\sqrt{1-z^2}}+
\int_{1}^{1+\pi /r} dz \frac{1}{\sqrt{1-(z-\frac{\pi}{r})^2}}+\right.
\no \\
 & &\left.+\int_{\pi /r}^1 dz \left( \frac{1}{\sqrt{1-z^2}}-
\frac{1}{\sqrt{1-(z-\frac{\pi}{r})^2}} \right) \right] = \no \\
 & &=\arcsin \sqrt{\frac{2\pi}{r}-\frac{\pi^2}{r^2}}.
\ea
Thus the integral of eq. (\ref{intcos}) is bounded by a
function with asymptotic behaviour ${\cal O}(1/\sqrt{r})$ and
lemma 1 follows. \hfill $\Box$

\paragraph{Lemma 2:} Let $f$ be a three times continuously
differentiable function and $w$ be a real number with $|w|<1$.
Then the integral
\be
I_w(r) \equiv \int_0^\pi d\theta \, \sin \theta \, \e^{-ir\cos
\theta } \sin (wr\cos \theta ) f(\theta )
\ee
has the asymptotic behaviour
\be
I_w(r) = \frac{\e^{-ir}}{r} f(0) \frac{w\cos wr + i\sin
wr}{1-w^2} - \frac{\e^{ir}}{r} f(\pi) \frac{w\cos wr - i\sin
wr}{1-w^2}  + {\cal O}(r^{-3/2})
\ee
in the limit $r \rightarrow \infty$.
\paragraph{Proof:} The proof of this lemma follows from lemma 1 by using
$\sin \alpha = (\e^{i\alpha}-\e^{-i\alpha})/2i$ for $\alpha =
wr\cos \theta$. \hfill $\Box$

\paragraph{Lemma 3:} Let $\psi : {\bf R}^3 \rightarrow {\bf
R}^3$ be a twice continuously differentiable function such that the
function itself and its first and second derivatives are
absolutely integrable. Then the integral
\be
I(\vq ) \equiv \int d^3q \, \psi(\vq ) \, \e^{-i\vq \cdot \vL}
\ee
decreases like $L^{-2}$ for $L \rightarrow \infty$.
\paragraph{Proof:} According to the assumption, $\triangle \psi$
is absolutely integrable and with a twofold partial integration
we obtain
\be
\int d^3q \, (\triangle \psi)(\vq) \, \e^{-i\vq \cdot \vL}=
-\vL^2 \int d^3q \, \psi(\vq) \, \e^{-i\vq \cdot \vL}
\ee
which proves lemma 3. \hfill $\Box$

\paragraph{Proof of the theorem:} The main point in the proof is to
meticulously take care to perform the limit $\epsilon
\rightarrow 0$ before investigating the asymptotic limit $L
\rightarrow \infty$.

\paragraph{Step 1:} For $A<0$ we apply lemma 3 and thus obtain
the second part of the theorem. In the following $A$ will always
be positive.

\paragraph{Step 2:} Next we split $J(\vL )$ into the integrals
\be
J_1(\vL ) = -\frac{i\pi}{2} \sa \int_{S^2} d\Omega \, \Phi(\sa \vn)
\, \e^{-i\sa \vn \cdot \vL}
\ee
originating in the $\delta -$function of the limit $\epsilon
\rightarrow 0$ and
\be
J_2(\vL ) = - \int d^3q \, \Phi(\vq) \, \e^{-i\vq \cdot \vL}
\frac{q^2-A}{(q^2-A)^2 + \epsilon^2}
\ee
which represents the principal value of this limit. $S^2$ denotes
the 2--dimensional unit sphere, $q \equiv |\vq |$ and $\vn
\equiv \vq/q$.

\paragraph{Step 3:} We choose two numbers $\delta$ and $\eta$
such that $0< \delta < \eta < \sa$ and a symmetric $C^\infty$
function $h$ such that $0 \leq h(v) \leq 1$ for all $v \in {\bf
R}$, $h(v)=1$ for $|v| \leq \delta$ and $h(v)=0$ for $|v| \geq
\eta$. Then $J_2 = \sum_{k=1}^3 J_{2k}$ with
\ba
J_{21}(\vL) & \equiv & -\int d^3q \, \e^{-i\vq \cdot \vL} \left(\Phi(\vq) -
\Phi(\sa \vn)h(q-\sa)\right) \frac{q^2-A}{(q^2-A)^2 +
\epsilon^2},
\\
J_{22}(\vL) & \equiv & -\int_{S^2} d\Omega \int_0^\infty dq \,
\e^{-i\vq \cdot \vL} \, \Phi(\sa \vn) \, h(q-\sa) \cdot \no \\
 & & \cdot \left(
\frac{q^2(q^2-A)}{(q^2-A)^2 +
\epsilon^2} - \frac{2A^{3/2} (q-\sa)}{4A(q-\sa)^2 + \epsilon^2} \right),
\\
J_{23}(\vL) & \equiv & -\int_{S^2} d\Omega \int_0^\infty dq \,
\e^{-i\vq \cdot \vL} \, \Phi(\sa \vn) \, h(q-\sa) \,
\frac{2A^{3/2} (q-\sa)}{4A(q-\sa)^2 + \epsilon^2}.
\ea
In these three integrals the limit $\epsilon \rightarrow
0$ can be performed. In $J_{21}$ one obtains an integrand which
is twice continuously differentiable and therefore $J_{21}$
decreases like $L^{-2}$ for $L \rightarrow \infty$ according to
lemma 3. The function within the parentheses of $J_{22}$ gives
$(q+\sa /2)/(q+\sa)$ in the limit $\epsilon \rightarrow 0$ and
with the properties of $h$ the same asymptotic behaviour as that
of $J_{21}$ results. Finally, with the transformation $v=q-\sa$
the integral $J_{23}$ can be cast into the form
\be
J_{23}(\vL)=\frac{i}{2} \sa \int_{S^2} d\Omega \, \Phi(\sa \vn) \, \e^{-i\sa
\vn \cdot \vL } \int_{-\eta}^\eta dv \, h(v) \, \frac{\sin (v\vn \cdot
\vL )}{v}.
\ee

\paragraph{Step 4:} To perform the integration over $S^2$ we take
\be
\vL = L \left( \begin{array}{c} 0 \\ 0 \\ 1 \end{array} \right) \equiv L \ve
\quad \mbox{and} \quad
\vn(\theta,\phi) = \left(
\begin{array}{c} \sin \theta \cos \phi \\ \sin \theta \sin \phi \\
\cos \theta \end{array} \right).
\ee
Then the application of lemma 1 with $r=\sa L$ shows that
\be
J_1(\vL)=\frac{\pi^2}{L} \left( \e^{-i\sa  L} \Phi(\sa \ve) -
\e^{i\sa  L} \Phi(-\sa \ve) \right) + {\cal O}(L^{-3/2}).
\ee

\paragraph{Step 5:} By virtue of lemma 2 and taking $w=v/\sa$, the
asymptotic behaviour of $J_{23}$ is given by
\ba
J_{23}(\vL) & = & \frac{i\pi}{L} \int_{-\eta}^\eta dv
\frac{h(v)}{v(1-\frac{v^2}{A})}
\left[ \e^{-i\sa  L} \Phi(\sa \ve)
\left( \frac{v}{\sa} \cos (vL) + i\sin (vL) \right) \right. - \no \\
 & & - \e^{i\sa  L} \Phi(-\sa \ve) \left.
\left( \frac{v}{\sa} \cos (vL)-i\sin (vL) \right) \right]
+ {\cal O}(L^{-3/2}).
\ea
The terms with $\cos (vL)$ decrease faster than any power of
$1/L$ by a corollary of lemma 3 whereas the $\delta-$function
property of $\sin (vL)/\pi v$ leads to
\be
J_{23}(\vL)= -\frac{\pi^2}{L} \left( \e^{-i\sa  L} \Phi(\sa \ve)+
\e^{i\sa  L} \Phi(-\sa \ve) \right) + {\cal O}(L^{-3/2}).
\ee
The correction to the $\delta -$function is proportional to
\be
\int\limits_{-\infty}^\infty dv \left( \frac{h(v)}{1-\frac{v^2}{A}} - 1
\right) \frac{\sin (vL)}{v}
\ee
and therefore decreases faster than any power of $1/L$ as can
easily be seen by repeated partial integration. Finally, the
theorem is obtained by summing the results of steps 4 and 5.
\hfill $\Box$

\end{document}